\documentclass[reprint]{revtex4-1}

\usepackage{graphicx}
\usepackage{dcolumn}
\usepackage{bm}
\usepackage{mathrsfs}

\begin{document}

\title{Limitations of minimum beam emittance at operational intensity in fourth-generation synchrotron light sources}

\author{Victor Smaluk}
\email{smaluk@bnl.gov} 
\affiliation{Brookhaven National Laboratory, Upton, NY 11973, USA}
\author{Timur Shaftan}
\affiliation{Brookhaven National Laboratory, Upton, NY 11973, USA}

\date{\today}

\begin{abstract}
For synchrotron light sources, the brightness of user X-ray beams is primarily determined by the electron beam emittance and energy spread at operational intensity. A common feature of fourth-generation synchrotrons is the short length of electron bunches combined with a very small transverse beam size. Consequently, the particle density is much higher than in machines of previous generations, leading to strong collective effects that significantly increase the emittance and limit the achievable brightness at operational beam intensity. In this article, we summarize our studies of the emittance scaled with the beam energy and intensity, taking into account the effects of intrabeam scattering, beam-impedance interaction, and bunch lengthening provided by higher-harmonic RF systems, to identify optimal combinations of machine and beam parameters.
\end{abstract}


\maketitle


\section{Introduction}\label{section1}

Over the past three decades, there has been a remarkable evolution in synchrotron light sources, transforming the landscape of scientific research and technological advancements. The ultimate brightness of light sources is the key to advancing to a smaller scale, faster response, and higher rate of data measurement and processing. The evolution of synchrotron light sources in past decades follows a path of continued increased brightness of photon beams. The development of third- and fourth-generation synchrotrons has enabled scientists to explore more complex and dynamic systems at unprecedented resolutions, from materials science and chemistry to biology and medicine. Looking ahead, ongoing advancements in accelerator technology and facility upgrades continue to shape the next generation of synchrotron light sources, promising even greater scientific insights and technological innovations.

Since brightness is one of the main figures of merit for synchrotron light sources, all projects of future synchrotrons consider an increase of brightness by a few orders of magnitude compared to the present standards. Brightness is a function of the electron beam energy, intensity, emittance, energy spread, and choice of light-generating insertion devices (wigglers and undulators). Reducing the emittance is a straightforward and efficient way to increase the brightness.

\section{Evolution of beam emittance}\label{section2}

The natural emittance of an electron beam in a storage ring is determined by a balance of the radiation damping and quantum excitation. For a planar ring without vertical bending, this relation is expressed as a ratio of radiation integrals~\cite{Helm-PAC1973} in the following equation
\begin{equation}
\varepsilon_{x0} = C_q\gamma^2 \frac{I_{5}}{I_2-I_{4}} \ .
\label{eq:emittance}
\end{equation}
Here $\varepsilon_{x0}$ is the emittance; $\gamma$ is the Lorentz factor; \mbox{$C_q = \frac{55}{32\sqrt{3}} \frac{\hbar c}{E_e} \simeq 3.83\cdot{10}^{-13}$~m}; the radiation integrals are
\begin{equation}
I_2=\!\!\int\!\!\frac{1}{\rho^2}ds \,, \quad I_4=\!\!\int\!\!\frac{\eta_x}{\rho^3}\left(1+2\rho^2K_1\right)ds \,, \quad I_5=\!\!\int\!\!\frac{\mathscr{H}}{|\rho|^3}ds\,;
\label{eq:I2I4I5}
\end{equation}
\begin{equation}
\mathscr{H}=\beta_x\eta_x'^2+2\alpha_x\eta_x\eta_x'+\gamma_x\eta_x^2\ ;
\label{eq:curlyH}
\end{equation}
$\beta_x$ is the amplitude function of betatron oscillation (beta function), $\alpha_x\equiv-\beta_x'/2$, $\gamma_x\equiv\frac{1+\alpha_x^2}{\beta_x}$; $\eta_x$ and $\eta_x'$ is the dispersion function and its derivative, respectively; $\rho$ is the local bending radius; \mbox{$K_1=\frac{1}{B\rho}\frac{\partial{B_y}}{\partial{x}}$} is the normalized quadrupole strength.

The emittance can be represented in a simple way as 
\begin{equation}
\varepsilon_{x0} = F\frac{E^2}{J_xN_B^3} \ ,
\label{eq:emittance_B}
\end{equation}
where $F$ is some function of the magnet lattice, $E$ is the electron energy, $J_x$ is the horizontal damping partition, and $N_B$ is the number of bending (dipole) magnets in the ring. 

\begin{figure}[!h]
   \centering
   \includegraphics*[width=1.0\columnwidth]{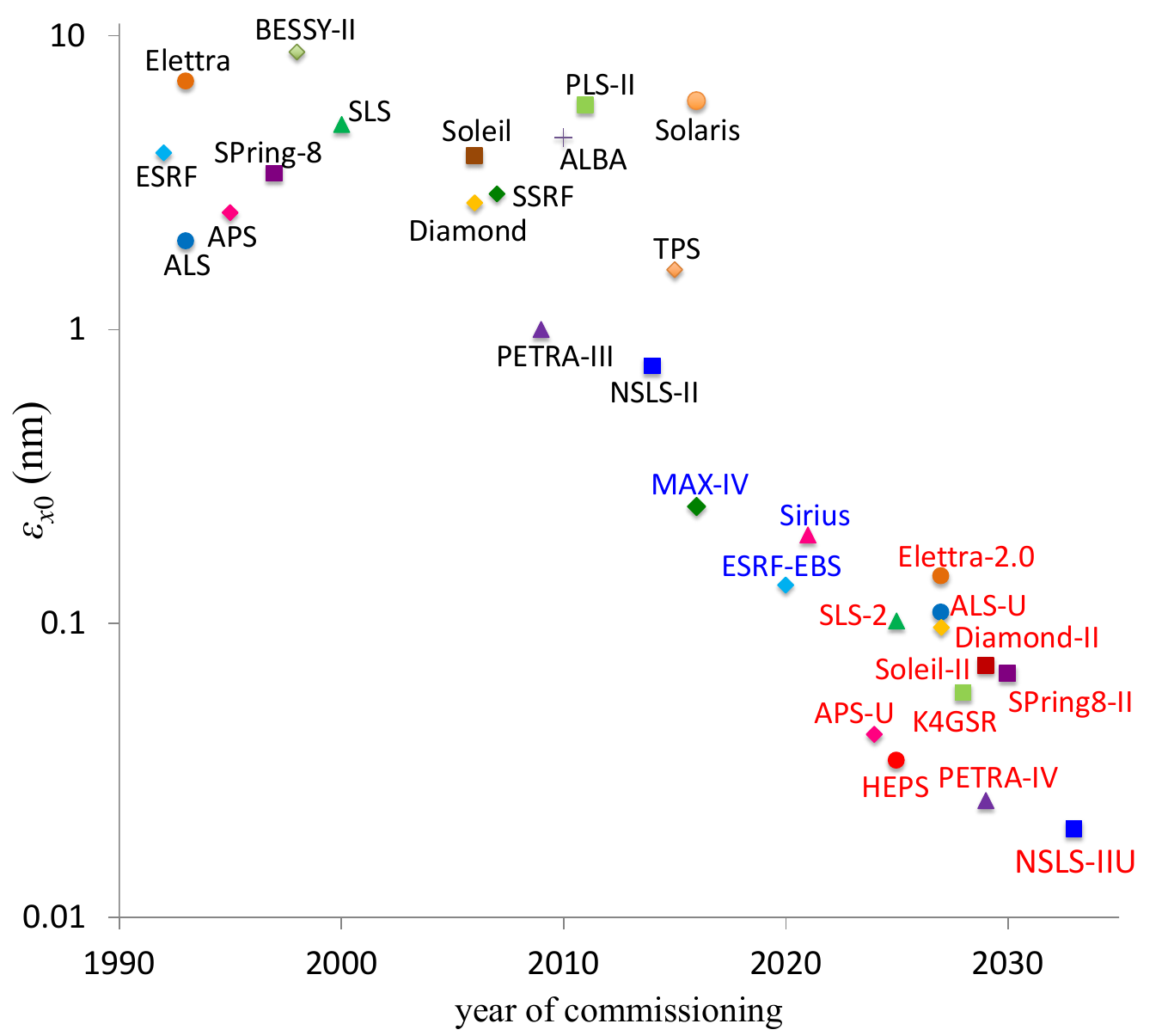}\\
   \caption{Evolution of the electron beam emittance in synchrotron light sources.}
   \label{fig:fig01}
\end{figure}

Since the emittance is inversely proportional to the cube of the number of bending magnets, increase of their number is a most efficient way to design a low-emittance lattice. As a result, we see transition from Double-Bend and Triple-Bend Achromat lattices  used to build 3rd-generation light sources worldwide, to Multi-Bend Achromat (MBA), which is the basic lattice option for new and upgrade projects of light sources. The fast development of advanced MBA technology resulted in a new generation of light source facilities. Recently, three new low-emittance synchrotrons based on MBA have been commissioned: MAX-IV (Sweden, 2016)~\cite{MAX4-2018}, ESRF-EBS (France, 2020)~\cite{ESRF-2023}, and SIRIUS (Brazil, 2020)~\cite{SIRIUS-2020}. One more project, APS-U~\cite{APS-U-2018} (USA), is in the commissioning stage, and HEPS~\cite{HEPS-2018} (China) will approach the commissioning soon. Many other projects of new and upgraded facilities are being developed worldwide, e.g. ALS-U~\cite{ALS-U-2017} (USA), Elettra-2~\cite{Elettra2-2018} (Italy), Diamond-II~\cite{Diamond2-2019} (UK), Soleil-2~\cite{Soleil2-2018} (France), PETRA IV~\cite{PETRA4} (Germany), CLS-2~\cite{CLS-2} (Canada), Korea-4GSR~\cite{PAL4GSR}. The beam emittance is continuously reduced in few decades of synchrotron development, as illustrated by Figure~\ref{fig:fig01}. 

There is a recent trend in magnet design towards combined magnets with field profiles tailored to the lattice requirements. In near future, we expect a transition to the permanent-magnet bending/focusing elements providing high quadrupole gradients, saving space, and reducing total power consumption. The use of superconducting high-field high-gradient magnets looks also promising for future projects.

A new approach of low-emittance lattice design alternative to MBA has been recently proposed at NSLS-II, Brookhaven National Laboratory: use of a new lattice element ``complex bend” replacing regular dipole magnets~\cite{BNL-211211-2019, PRAB-21-100703, BNL-211223-2019, PRAB-2019-110703}. The main advantage of the complex bend design is to enable many more dense dipoles integrated into the same element. Since the emittance scales inversely as the cube of the number of dipoles, this opens a possibility of achieving gains in emittance. For example, replacement of the dipole magnets with complex bends in the NSLS-II DBA lattice keeping the layout of matching quadrupole triplets and straight sections unchanged, results in the emittance reduction by a factor of 30~\cite{JPhys-1350-012044}. More advanced options of the complex bend lattice design for the NSLS-II upgrade assuming the replacement of the whole ring provide even lower emittance of about 24~pm~\cite{PRAB-2021-HCBA,NSLS2U-CBA}. 

\section{Energy and intensity constraints}\label{section3}

The major practical limit of the electron beam current $I_b$ and energy $E$ in a storage ring results from the synchrotron radiation power rapidly increasing with the beam energy as $E^4$:
\begin{equation}
P_\mathrm{rad} = I_bU_0 = I_b \frac{C_\gamma}{2\pi} E^4 I_2 \ , 
\label{eq:Prad}
\end{equation}
where $U_0$ is the energy loss per turn, \mbox{$C_\gamma = 8.846\cdot 10^{-5}$~m/GeV$^3$}, $I_2$ is the 2-nd radiation integral \mbox{$I_2=\oint\frac{ds}{\rho}$}, and $\rho$ is the local radius of electron trajectory curvature. 

In modern synchrotrons, the total radiation power is dominated by the light-generating insertion devices (wigglers and undulators), whose contribution is usually higher than the contribution of dipole magnets. The beam energy loss caused by radiation is compensated by complex and expensive RF systems, which contribute a significant part to the total cost of the machine construction and operation, due to high power consumption. 

\begin{figure}[!h]
   \centering
   \includegraphics*[width=1.0\columnwidth]{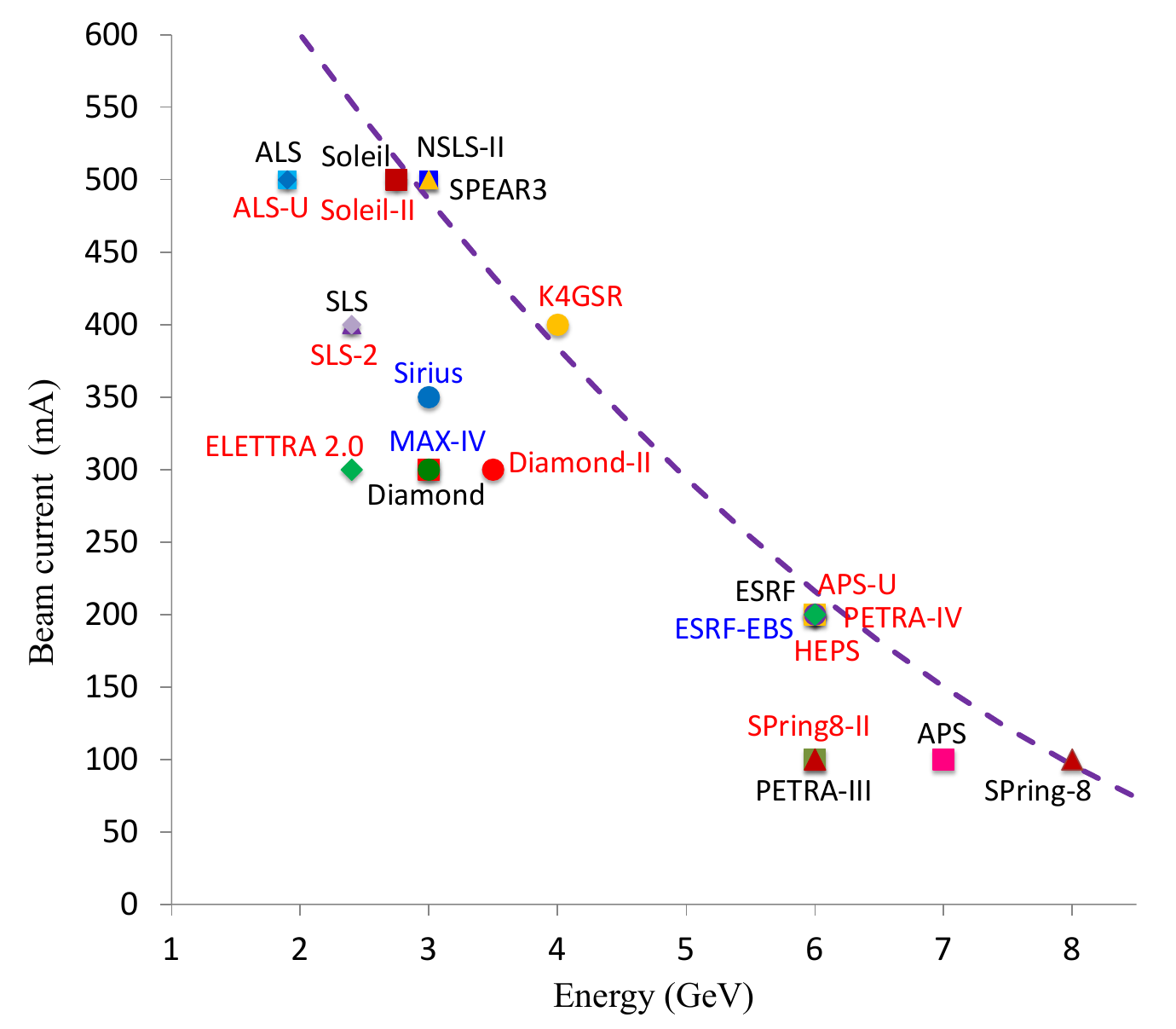}\\
   \caption{Operation beam current of synchrotrons vs beam energy.}
   \label{fig:fig02}
\end{figure}

Mainly for the above-mentioned reason, higher-energy synchrotrons operate with lower beam intensity, as illustrated in Figure~\ref{fig:fig02}. The dashed line represents an empirical limit curve for the beam intensity.

\section{Intensity-dependent effects}\label{section4}

Ideally, the photon flux and brightness should be directly proportional to the electron beam current. However, collective effects of beam dynamics at operational beam intensity play a crucial role in determining the practically achievable performance of light sources. In modern low-emittance rings, electron beams are small in all three dimensions: a small momentum compaction results in a short bunch length, while a low emittance determines small transverse sizes. 

\begin{figure}[!h]
   \centering
   \includegraphics*[width=1.0\columnwidth]{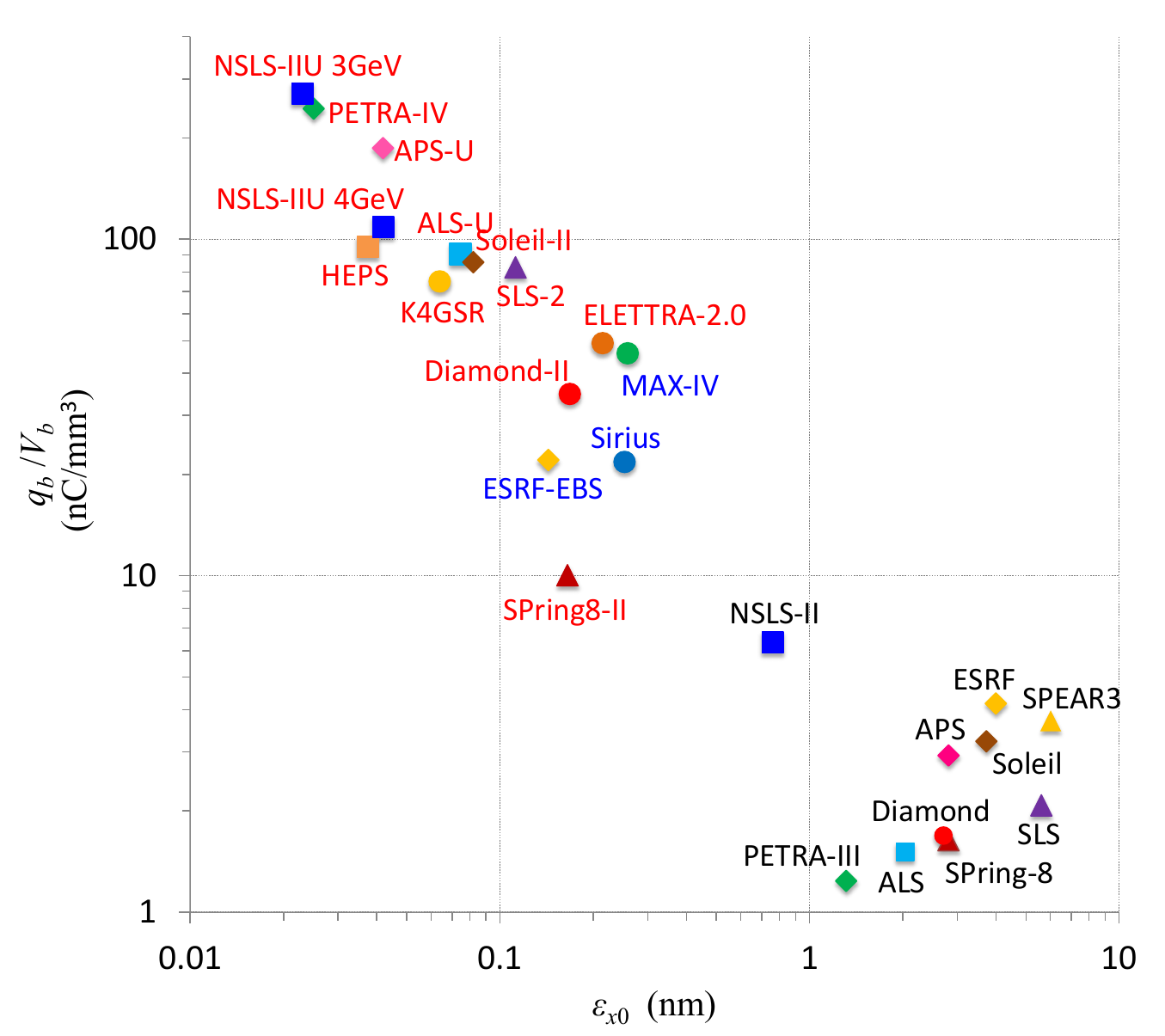}\\
   \caption{Charge density in synchrotrons vs beam emittance.}
   \label{fig:fig03}
\end{figure}

Figure~\ref{fig:fig03} demonstrates the charge density $q_b/V_b$ 
as a function of the emittance for a set of synchrotrons worldwide, in operation or under development. A trend of a significant increase of the particle density within the bunch in modern low-emittance synchrotrons is clearly seen resulting in much stronger collective effects. Here \mbox{$V_b = \sqrt{4\pi^3} \sigma_x \sigma_y \sigma_z$} is the bunch volume, $\sigma_z$ is the r.m.s.\ bunch length, $\sigma_{x,y}$ is the horizontal/vertical beam size. Note that the bunch volume was calculated at zero beam intensity and without higher-harmonic cavities widely used to increase the bunch length for mitigation of collective effects, so the bunch length $\sigma_z$ is completely determined by the lattice and RF parameters:

\begin{equation}
\sigma_z = \sigma_\delta \sqrt{\frac{\lambda_\mathrm{RF}R_\mathrm{aver}\alpha_c E}{\sqrt{e^2V_\mathrm{RF}^2-U_0^2}}} \ , 
\label{eq:sigmaz}
\end{equation}
where $\sigma_\delta$ is the relative energy spread, $\lambda_\mathrm{RF}$ is the RF wavelength, $V_\mathrm{RF}$ is the RF voltage, $R_\mathrm{aver}$ is the average ring radius, $\alpha_c$ is the momentum compaction factor.

Intra-beam scattering (IBS) is one of the adverse effects that can impact beam quality and impose limitations on the ultimate performance of low- and medium-energy synchrotrons~\cite{ALSU-IBS, NSLS2-IBS}. Since IBS is a small-angle scattering, it does not cause particle loss but results in a substantial intensity-dependent increase in emittance, energy spread, and bunch length. The theory of IBS has been well-developed for quite some time~\cite{IBS-Piwinski, IBS-Bjorken, IBS-Bane, IBS-Kubo} and has been implemented into particle tracking codes~\cite{elegant}.

We employed the high-energy approximation of the IBS theory~\cite{IBS-SLAC-ATF} to examine the combined effect of IBS and the bunch lengthening resulting from the longitudinal impedance and higher-harmonic cavities. The equilibrium emittance \mbox{$\varepsilon_{x,y}$} and relative energy spread \mbox{$\sigma_\delta$} are expressed as follows:
\begin{equation}
\varepsilon_{x,y} = \frac{\varepsilon_{x0,y0}}{1-\tau_{x,y}/T_{x,y}}\ ,  \qquad  \sigma_\delta^2 = \frac{\sigma_{p0}^2}{1-\tau_p/T_p} \ ,
\label{eq:IBS_emit}
\end{equation}
where $\varepsilon_{x0,y0}$ and $\sigma_{p0}$ are the emittance and energy spread at zero beam current; $\tau_x$, $\tau_y$, and $\tau_p$ are the radiation damping times; $T_x$, $T_y$, and $T_p$ are the IBS growth times:
\begin{equation}
\frac{1}{T_p} \simeq \frac{r_0^2cN}{32\gamma^3\varepsilon_x\varepsilon_y\sigma_z\sigma_\delta^2} \left(\frac{\varepsilon_x\varepsilon_y}{\left<\beta_x\right>\left<\beta_y\right>}\right)^{\!\!1/4} \ln\frac{\left<\sigma_y\right>\gamma^2\varepsilon_x}{r_0\left<\beta_x\right>} \ ,
\label{eq:IBS_Tp}
\end{equation}
\begin{equation}
\frac{1}{T_{x,y}} \simeq \frac{\sigma_\delta^2\left<\mathscr{H}_{x,y}\right>}{\varepsilon_{x,y}} \frac{1}{T_p} \ ,
\label{eq:IBS_Txy}
\end{equation}
$r_0$ is classical electron radius, $\mathscr{H}_{x,y}$ is a function determined by the lattice~(\ref{eq:curlyH}).

Since the IBS strongly depends on the beam energy, its effect is not so significant for high-energy rings.



In a practical range of the beam energy and current, we analyzed the impact of intensity-dependent effects on the light source performance using the complex bend lattice for NSLS-II upgrade optimized to achieve a minimum emittance, decent dynamic aperture, and beam lifetime~\cite{NSLS2U-CBA}. 

We calculated the emittance as a function of the beam current and energy for this lattice, considering both IBS and impedance effects. The bunch lengthening caused by the beam interaction with the longitudinal impedance was computed using the modified Zotter equation~\cite{zotter1981,zhow2023}:
\begin{equation}
\bigg(\frac{\sigma_t}{\sigma_{t0}}\bigg)^{\!3} -\frac{\sigma_t}{\,\sigma_{t0}} = \frac{I_b\,\alpha_c}{4\!\sqrt{\pi}\,\nu_s^2\,\omega_0^3\sigma_{t0}^3\,E/e}\,
\mathrm{Im}\left(\frac{Z_\parallel}{n}\right)_\mathrm{eff} \ ,
\label{eq:bunch_lengthening_potential_well}
\end{equation}
where \mbox{$\nu_s=\omega_s/\omega_0$} is the synchrotron tune, $\sigma_{t0}$ is the bunch length at zero intensity, $(\mathrm{Im}Z_\parallel/n)_\mathrm{eff}$ is the effective normalized longitudinal impedance.

\begin{figure}[!h]
   \centering
   \includegraphics*[width=1.00\columnwidth]{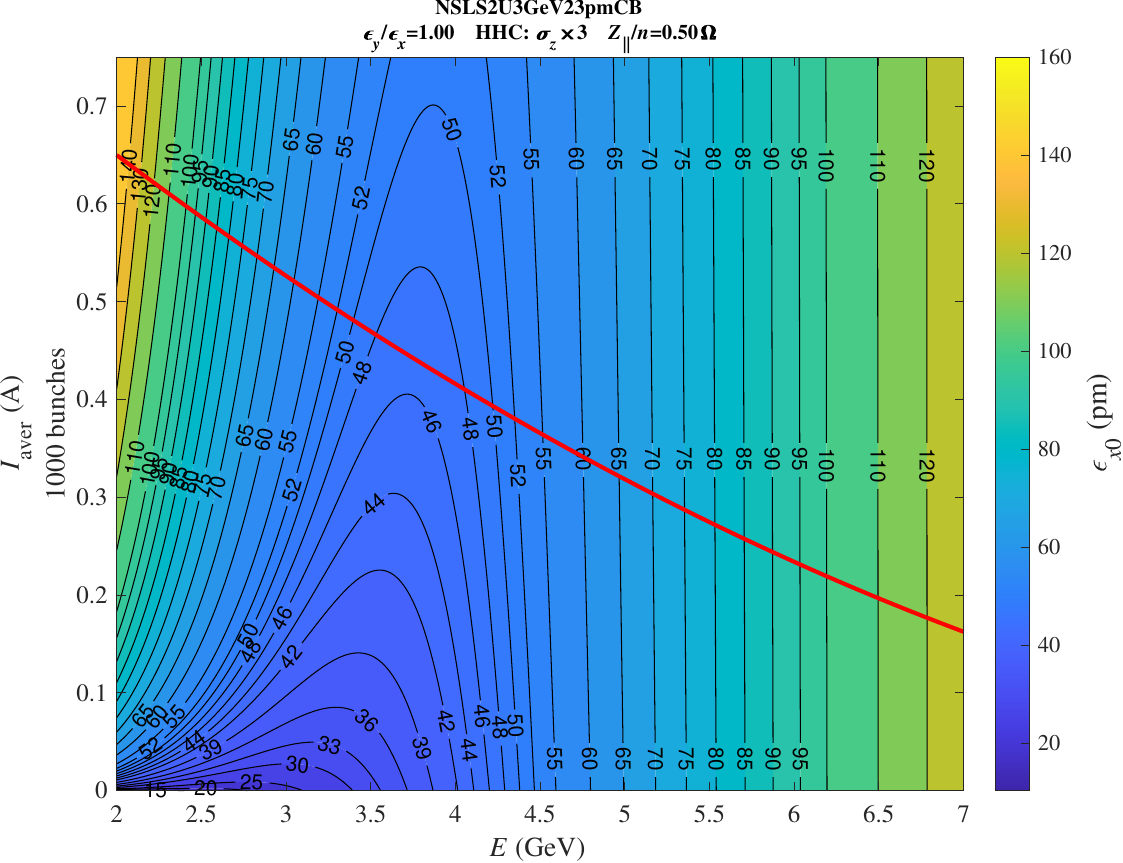}\\
   \caption{Combined effect of IBS, impedance, and higher-harmonic cavities on the beam emittance.}
   \label{fig:fig04}
\end{figure}

The RF voltage was scaled with the energy to keep the RF acceptance of 5\%. The effect of higher-harmonic cavities was simply modeled by multiplying the zero-intensity bunch length $\sigma_{t0}$ by a moderate factor of~3. In this calculation, we assumed the "bare" lattice without wigglers and undulators, 100\% coupling ($\varepsilon_y=\varepsilon_x)$, and a typical normalized longitudinal impedance $\mathrm{Im}\left(Z_\parallel/n\right)_\mathrm{eff}=0.5~\Omega$. Figure~\ref{fig:fig04} shows the emittance $\varepsilon_{x0}$ affected by these collective effects, as a function of the energy $E$ and average beam current $I_\mathrm{aver}$ uniformly distributed in 1000 bunches.

We found that the emittance at the operational beam intensity reaches a minimum in the energy range of \mbox{3.5-4~GeV}. This is due to the strong emittance blow-up caused by IBS at lower energies, while the $E^2$ factor in equation (\ref{eq:emittance_B}) leads to an increase in emittance at higher energies. So there is an optimal energy with the smallest emittance at the operational beam intensity for any particular lattice.

The light-generating insertion devices (IDs) contribute significantly to the total energy loss per turn $U_0$ determining the radiation damping, emittance, and energy spread. Thus for the complex bend lattice~\cite{NSLS2U-CBA} with a full set of IDs, the total energy loss is about 1~MeV compared to 0.3~MeV of the radiation energy loss from bending magnets only. 

\begin{figure}[!h]
   \centering
   \includegraphics*[width=1.00\columnwidth]{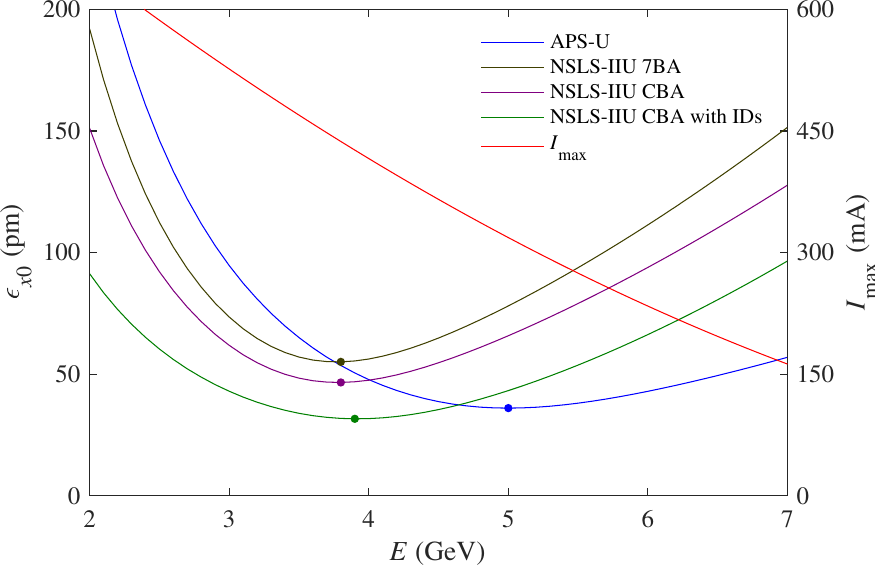}\\
   \caption{Minimum emittance along the the energy-dependent operational beam current.}
   \label{fig:fig05}
\end{figure}

Assuming the empirical energy-dependent limit of the operational beam current presented as the dashed line in Figure~\ref{fig:fig02}, we carried out emittance scaling with the energy for the following lattices: 1) a hybrid 7BA lattice for the APS Upgrade~\cite{APS-U-2018}, which is the most elaborated MBA lattice in terms of emittance minimization; 2) an MBA lattice based on the ESRF-EBS design scaled to the size of NSLS-II and optimized to achieve minimum emittance~\cite{NSLS2U-H7BA}; 3) the complex bend lattice for NSLS-II upgrade without IDs; 4) the complex bend lattice with a full set of IDs. The results are presented in Figure~\ref{fig:fig05}, the minimum emittance at the operational beam current indicates the optimal energy for each lattice. Note, the insertion devices slightly move the optimal energy up.

Note the bunch volume in the denominator of the IBS growth rate~(\ref{eq:IBS_Tp}) is also relevant for the Touschek lifetime, which is also a practical intensity-limiting effect in 4th-generation synchrotrons. Since lower emittance needs stronger beam focusing resulting in high natural chromaticity, strong sextupole magnets are needed to compensate for it. So beam dynamics becomes very nonlinear with smaller momentum acceptance leading to the degradation of the Touschek lifetime. While the DBA and TBA lattices of 3rd-generation synchrotrons typically operate with a beam current up to 500~mA and a beam lifetime exceeding 10 hours, it is about an order of magnitude lower for most of the new MBA-based machines. 

Another important factor that could limit the beam intensity is the beam-induced heating of vacuum chambers, which is directly proportional to the longitudinal impedance (mainly the resistive-wall one) and to the square of the beam current. Strong focusing magnets and high-brightness insertion devices require low-aperture vacuum chambers. Since the longitudinal impedance is inversely proportional to the vacuum chamber size, small vacuum chambers and short bunch lengths lead to higher beam-induced power. Moreover, the transverse impedance is inversely proportional to the cube of the vacuum chamber size. The beam interaction with the impedance can also lead to beam instabilities, further limiting the maximum stable beam current. The interaction with the residual gas in vacuum chambers can excite the instabilities too, and this problem is also more severe for the modern synchrotrons because the vacuum chambers are smaller and the pumping is more difficult.

There are several ways to mitigate the challenges caused by collective effects. Lattice optimization can help to increase natural bunch length. To reduce the peak current of the beam, bunch lengthening is essential and can be achieved by implementing advanced schemes of higher-harmonic cavities. Increasing number of bunches is also helpful to reduce the peak current and beam-induced heating of the vacuum chamber. To reduce the impedance, larger vacuum chamber should be used where possible and smooth transitions must be implemented. Minimization of the impedance needs to be a part of the vacuum chamber design from the very beginning of a project. 

\section{Conclusion}\label{conclusion}

Next-generation synchrotrons have a common feature of short electron bunches and a small transverse beam size, resulting in a significant reduction in bunch volume, higher particle density, and stronger collective effects. Since the brightness of user X-ray beams at operational intensity is predominantly determined by the electron beam emittance, we studied the intensity-dependent emittance scaled with beam energy, considering the effects of intrabeam scattering, beam-impedance interaction, and bunch lengthening by higher-harmonic RF cavities, to identify optimal combinations of machine and beam parameters. The emittance at operational beam intensity reaches a minimum at specific energy due to the interplay of the quadratic increase of zero-intensity emittance and IBS-induced blow-up at lower energies. This optimal energy point with the smallest operational emittance is determined by a specific lattice design and slightly affected by wigglers and undulators.

\section*{Acknowledgments}
This work was supported by the U.S. Department of Energy under Contract No. DE-SC0012704 and by Brookhaven National Laboratory Directed Research and Development Program, Project No. 20-041.


\begin{thebibliography}{99}

\bibitem{Helm-PAC1973} 
R.H.~Helm, M.J.~Lee, P.L.~Morton, M.~Sands, ``Evaluation of Synchrotron Radiation Integrals", in Proc. of PAC-1973, San Francisco, pp.~900-901.

\bibitem{MAX4-2018}
P.~Tavares, E.~Al-Dmour, A.~Andersson, F.~Cullinan, B.~Jensen, D.~Olsson, D.K.~Olsson, M.~Sjostrom, H.~Tarawneh, S.~Thorina, A.~Vorozhtsov, ``Commissioning and first-year operational results of the MAX IV 3 GeC ring", J. Synchrotron Rad. 25 (2018) 1291--1316.

\bibitem{ESRF-2023}
P.~Raimondi, et al., ``The Extremely Brilliant Source storage ring of the European Synchrotron Radiation Facility", Commun. Phys. 6, 82 (2023).

\bibitem{SIRIUS-2020}
L.~Lin, ``Sirius Accelerators Overview", in Proc. of EIC Workshop, 2020.

\bibitem{HEPS-2018}
Y.~Jiao, et al., ``The HEPS project", J. Synchrotron Rad. 25 (2018) 1611--1618.

\bibitem{APS-U-2018}
M.~Borland, et al., ``The Upgrade of the Advanced Photon Source", in Proc. of IPAC-2018, Vancouver, THXGBD1.

\bibitem{ALS-U-2017}
C.~Steier, et al., ``Status of the Conceptual Design of ALS-U", in Proc. of IPAC-2017, Copenhagen, WEPAB104.

\bibitem{Elettra2-2018} E. Karantzoulis, ``Elettra 2.0 - The diffraction limited successor of Elettra" , Nuclear Inst. and Methods in Physics Research, A 880 (2018) 158-165.

\bibitem{Diamond2-2019}
Diamond-II: Conceptual Design Report, 2019.

\bibitem{Soleil2-2018}
A.~Loulergue, ``Baseline Lattice for the Upgrade of {SOLEIL}", in Proc. of FLS-2018, Shanghai, MOP2WB03.

\bibitem{PETRA4}
C.G.~Schroer, H.C.~Wille, O.H.~Seeck, et al., `The synchrotron radiation source {PETRA III} and its future ultra-low-emittance upgrade {PETRA IV}", Eur. Phys. J. Plus 137, 1312 (2022). 

\bibitem{CLS-2}
L.O.~Dallin, ``Design Considerations for an Ultralow Emittance Storage Ring for the Canadian Light Source", in Proc. of IPAC-2018, Vancouver, TUPMF038.

\bibitem{PAL4GSR}
S.~Shin, ``Lattice Design and Beam Dynamics Studies for Fourth-Generation Storage Ring", 23rd International Conference on Accelerators and Beam Utilization, Daejeon, 2019.

\bibitem{BNL-211211-2019}
T.~Shaftan, V.~Smaluk, G.~Wang, ``A Concept of the Complex Bend", BNL-211211-2019-TECH, Upton, 2018.

\bibitem{PRAB-21-100703}
G.~Wang, T.~Shaftan, V.~Smaluk, N.~Mezentsev, S.~Sharma, O.~Chubar, Y.~Hidaka, C.~Spataro, ``Complex bend: Strong-focusing magnet for low-emittance synchrotrons", Phys. Rev. Accel. Beams 21 (2018) 100703.

\bibitem{BNL-211223-2019}
T.~Shaftan, G.~Wang, V.~Smaluk, Y.~Hidaka, O.~Chubar, T.~Tanabe, J.~Choi, ``Complex Bend II", BNL-211223-2019-TECH, Upton, 2018.	

\bibitem{PRAB-2019-110703}
G.~Wang, T.~Shaftan, V.~Smaluk, Y.~Hidaka, O.~Chubar, T.~Tanabe, J.~Choi, S.~Sharma, C.~Spataro, N.~Mesentsev, ``Complex bend. II. A new optics solution",	 Phys. Rev. Accel. Beams 22 (2019) 110703.

\bibitem{JPhys-1350-012044}
V.~Smaluk, T.~Shaftan, ``Realizing low-emittance lattice solutions with Complex Bends", J. Phys.: Conf. Ser. 1350 (2019) 012044.

\bibitem{PRAB-2021-HCBA}
F.~Plassard, G.~Wang, T.~Shaftan, V.~Smaluk, Y.~Li, Y.~Hidaka, ``Simultaneous correction of high order geometrical driving terms with octupoles in synchrotron light sources", Phys. Rev. Acc. Beams 24 (2021) 114801.

\bibitem{NSLS2U-CBA}
M.~Song, T.~Shaftan, ``Design study of a low emittance complex bend achromat lattice", arXiv:2310.20010v2 (2023).

\bibitem{ALSU-IBS}
C.~Steier, J.~Byrd, H.~Nishimura, D.~Robin, S.~De~Santis, F.~Sannibale, C.~Sun, M.~Venturini, W.~Wan, ``Physics Design Progress Towards a Diffraction Limited Upgrade of the ALS", in Proc. of IPAC-2016, Busan, WEPOW049.

\bibitem{NSLS2-IBS}
A.~Blednykh, B.~Bacha, G.~Bassi, V.~Smaluk, T.~Shaftan, M.~Borland, R.~Lindberg, ``Combined Effect of IBS and Impedance on the Longitudinal Beam Dynamics", in Proc. of IPAC-2021, Campinas, THPAB240.

\bibitem{IBS-Piwinski}
A.~Piwinski, ``Intra-beam Scattering", in Proc. of 9th Int. Conf. on High Energy Accelerators, Stanford, 1974, p. 405.

\bibitem{IBS-Bjorken}
J.D.~Bjorken, S.K.~Mtingwa, ``Intrabeam Scattering",  Particle Accelerators  Vol. 13 (1983) pp. 115-143.

\bibitem{IBS-Bane}
K.~Bane, ``A Simplified Model of Intrabeam Scattering", in Proc. of EPAC-2002, Paris, WEPRI120.

\bibitem{IBS-Kubo}
K.~Kubo, S.K.~Mtingwa, A.~Wolski, ``Intrabeam scattering formulas for high energy beams", Phys. Rev. ST Accel. Beams 8 (2005) 081001.

\bibitem{elegant}
M.~Borland, ``elegant: A Flexible SDDS-Compliant Code for Accelerator Simulation", Advanced Photon Source LS-287, September 2000.

\bibitem{IBS-SLAC-ATF}
K.~Bane, H.~Hayano, K.~Kubo, T.~Naito, T.~Okugi, J.~Urakawa, ``Intrabeam Scattering Analysis of ATF Beam Measurements", SLAC-PUB-8875 (2001).

\bibitem{NSLS2U-H7BA}
Y.~Li, K.~Hwang, C.~Mitchell, R.~Rainer, R.~Ryne, V.~Smaluk,	 ``Design of double- and multi-bend achromat lattices with large dynamic aperture and approximate invariants", Phys. Rev. Accel. Beams 24 (2021) 124001.

\bibitem{zotter1981}
B.~Zotter, ``Potential-well bunch lengthening", CERN SPS/81-14 (DI). Geneva, Switzerland, 1981.

\bibitem{zhow2023}
D.~Zhou, G.~Mitsuka, T.~Ishibashi, K.~Bane, ``Potential-well Bunch Lengthening in Electron Storage Rings", arXiv:2309.00808.


\end{thebibliography}
\end{document}